\documentclass[]{aastex631}

\usepackage{amsmath}

\received{3 August, 2021}
\revised{10 October, 2021}
\accepted{13 October, 2021}

\submitjournal{ApJ}

\shorttitle{Density filling factor in the solar corona due to turbulence}
\shortauthors{Sen \& Pant}

\begin{document}

\title{How transverse MHD wave-driven turbulence influences the density filling factor in the solar corona?}

\author[0000-0003-1546-381X]{Samrat Sen$^\dagger$}
\author[0000-0002-6954-2276]{Vaibhav Pant$^{\ddagger}$}
\affiliation{Aryabhatta Research Institute of Observational Sciences,
Manora Peak, Nainital-263001, Uttarakhand, India\\ 
email: \url{samrat@aries.res.in}$^\dagger$, \url{samratseniitmadras@gmail.com}$^\dagger$, 
\url{vaibhav.pant@aries.res.in}$^\ddagger$}

\begin{abstract}

It is well established that the transverse MHD waves are ubiquitous in the solar corona. One of the possible mechanisms for heating both open (e.g. coronal holes) and closed (e.g. coronal loops) magnetic field regions of the solar corona is due to the MHD wave-driven turbulence. In this work, we have studied the variation in the filling factor of overdense structures in the solar corona due to the generation of the transverse MHD wave-driven turbulence. Using 3D MHD simulations, we estimate the density filling factor of an open magnetic structure by calculating the fraction of the volume occupied by the overdense plasma structures to the entire volume of the simulation domain. Next, we perform forward modeling and generate synthetic spectra of Fe XIII 10749 \AA\ and 10800 \AA\ density sensitive line pairs using FoMo. Using the synthetic images, we again estimate the filling factors. The estimated filling factors obtained from both methods are in reasonable agreement. Also, our results match fairly well with the observations of filling factors in coronal holes and loops. Our results show that the generation of turbulence increases the filling factor of the solar corona.

\end{abstract}


\section{Introduction} \label{sec:intro}

The study of the magnetic and thermodynamic structure of solar corona is an important aspect as it plays a key role in coronal heating up to millions of kelvin. There are mainly two schools of thought for the coronal heating problem: dissipation of magnetohydrodynamic (MHD) waves in the solar atmosphere, and magnetic reconnection. Evidence of MHD waves and their different modes have been reported for both ground and space-based observations in open magnetic field regions \cite[e.g.][]{ 2020arXiv201208802B}. The MHD waves can dissipate energy and heat solar corona as it propagates through the medium (\cite{1998A&A...337L...9R, 2006SoPh..234...41K, 2015RSPTA.37340261A, 2015NatCo...6.7813M}, and references therein). Due to the partial reflection of the high-frequency torsional Alfv{\'e}n waves in the transition region, a significant amount of energy flux ($\sim 10^3$ W m$^{-2}$) is transferred onto the corona \citep{2017NatSR...743147S}. The other mechanism of the coronal heating is the topological rearrangement of the magnetic field lines due to the motion of the magnetic footpoints leading to the magnetic reconnection which releases energy in the solar corona (\cite{1988ApJ...330..474P, 2005ESASP.596E..14P, 2013ApJ...769...59T}, and references therein). Along with the wave dissipation and reconnection, another class of coronal heating mechanism that has drawn attention in the solar community from the last decade is the wave-driven MHD turbulence \citep{1996ApJ...457L.113E, 1999ApJ...523L..93M, 2005ApJS..156..265C, 2008ApJ...677.1348R, 2011ApJ...736....3V, 2014ApJ...782...81V}. The wave-driven turbulence is either generated by the counter-propagating waves \citep{2001PhPl....8.2377D} or uni-directionally propagating transverse MHD waves in transversely inhomogeneous medium \citep[also called uniturbulence;][]{2017NatSR...714820M,2019ApJ...882...50M,2020ApJ...899..100V}.

The counter-propagating transverse MHD waves observed as the quasi-periodic Doppler velocity shifts were reported by \citet[][and references therein]{2007Sci...317.1192T, 2008SoPh..247..411T,2009ApJ...697.1384T, 2015NatCo...6.7813M} using COronal Multi-channel Polarimeter \citep[CoMP;][]{2008SoPh..247..411T}. The estimated energy carried by the transverse waves was found to be 2-3 orders of magnitude less than that required for heating the solar corona \citep{2007Sci...317.1192T}. This discrepancy could be due to the unresolved Doppler velocity shifts due to line of sight superposition of different oscillating structures \citep{2012ApJ...761..138M,2012ApJ...746...31D,2019ApJ...881...95P}. Thus the true wave energies are hidden in the non-thermal line widths \citep{2019ApJ...881...95P}. Therefore, a relation between the nonthermal line widths and rms wave amplitudes is given by \citet{2020ApJ...899....1P} to correctly estimate the wave energy carried by the bulk Alfv{\'e}n waves. However, using the relation of the energy density of the bulk Alfv{\'e}n waves overestimates the true energy \citep{2012ApJ...753..111G}. Also, \cite{2014ApJ...795...18V} has shown that the wave energies are reduced by several factors due to the inclusion of the filling factor. The study of the filling factors helps us to understand the overall density structure, and how the overdense regions are spatially distributed in the medium. This gives an estimation of the density inhomogeneity of the medium \citep{2016ApJ...829...42H}. On the other hand, kink waves (body waves) have maximum wave energy inside the overdense plasma structures, while the surface waves (Alfv{\'e}n waves) have energy concentrated at the boundaries of it \citep{2014ApJ...795...18V}. In this regard density inhomogeneity, and hence the study of the density filling factor is imperative to estimate the true energy flux carried by the waves in the solar corona.

We define the density filling factor as the fraction of the overdense volume which is embedded within a low-density region. A medium with the density filling factor close to unity indicates that the medium is nearly homogeneously filled with overdense structures. Filling factor in the open magnetic structures (e.g. coronal holes (CHs), coronal bright points, plumes, interplumes, moss regions) has been calculated in past \citep{2008ApJ...686L.131W, 2010A&A...518A..42T, 2016ApJ...829...42H}. These regions are the open magnetic structures, where the magnetic field lines rise but never come back in the same horizontal plane within the region of interest \citep{2019ApJ...877..127S}. \cite{2008A&A...491..561D} has estimated the filling factor of the coronal bright points in the quiet Sun (QS) regions by EUV Imaging Spectrometer (EIS) onboard \textit{Hinode} \citep{2007SoPh..243...19C} data, and found it to vary from $4\times 10^{-5}$ to 0.20 with a median value of 0.015. \cite{2014A&A...561A..29V} have calculated the filling factors varying from 0.02 to 0.22 in active regions and 0.02 to 0.18 in the quiet sun regions. \cite{2016ApJ...829...42H} have inferred the filling factors using ``Two-slab density model" in the coronal holes, plume, interplume, and quiet sun regions observed by EIS/Hinode data. They have found the filling factors to vary between 0.10 to 0.20 in the interplume regions, 0.004 to 0.02 in the plumes, and 0.10 in the CHs and QS regions. The volumetric filling factor of the coronal loops can be calculated by emission measure (EM), plasma density of the medium, and the apparent column depth. Apart from open magnetic field regions, \cite{2015ApJ...800..140G} have calculated the filling factor along an entire coronal loop observed by EIS/Hinode and Atmospheric Imaging Assembly \citep[AIA;][]{2012SoPh..275...17L} onboard \textit{Solar Dynamics Observatory} (SDO), and found it to vary from 0.11 at the footpoint and exceeds 1 at the loop top. Using the emission lines of EIS/Hinode, \cite{2010A&A...518A..42T} have estimated the filling factor for five different moss regions, which varies from 0.002 to 0.1 for Fe XII, 0.03 to 1.9 for Fe XIII, and 0.2 to 1.8 for Fe XIV lines. In spite of these observational advances in the estimations of the filling factors in the solar corona, our understanding of the physical mechanism responsible for the generation of filling factors and its variation with height is limited.

In this work, we infer the effects of transverse MHD wave-driven turbulence on the filling factor of the plasma using two different methods. First, by taking the ratio of the volume of overdense plasma columns with density more than a threshold value to the entire volume of the simulation domain. Second, we generate the synthetic intensity images of Fe XIII 10749 \AA\ and 10800 \AA\ density sensitive emission lines by forward modeling using FoMo. We compare the estimated filling factors obtained from both methods and with the observations in solar coronal structures. Also, we investigate the variation of the filling factor with height due to the presence of turbulence. The rest of the paper is organized as follows. In section \ref{sec: simulation}, we discuss the simulation setup and the input parameters of the simulation. In section \ref{sec:filling factor}, we describe the methods for estimating the filling factor from area estimation in \ref{subsec:filling factor from area}, and from forward modeling in \ref{subsec:filling factor-fomo}, and report the main findings of the work. In Section \ref{sec: discussion}, we compare our estimated values with the observations,  discuss the novelty of the work, and finally conclude our findings. 

\section{Simulation setup} \label{sec: simulation}

To understand the effect of the MHD wave-driven turbulence on the filling factor of an open field region of the solar corona (e.g. coronal holes), we use an ideal 3D MHD simulation using MPI-AMRVAC \citep{2014ApJS..214....4P}. A detailed description of the simulation setup is given in \cite{2019ApJ...881...95P}. The simulation has a grid size of 128 $\times$ 512 $\times$ 512 which spans 50 Mm $\times$ 5 Mm $\times$ 5 Mm in the spatial scale that solves the set of the following MHD equations:

\begin{align} \label{mhd1}
\frac{\partial \rho}{\partial t} + \nabla \cdot (\rho \textbf{v})&=0,\\ \label{mhd2}
\frac{\partial (\rho \textbf{v})}{\partial t}+\nabla (\rho \textbf{v}\textbf{v}-\textbf{B}\textbf{B}/\mu_0)+\nabla(p+B^2/2\mu_0)-\rho \textbf{g}&=0,\\ \label{mhd3}
\frac{\partial E}{\partial t}+\nabla \cdot[ (\textbf{v}E-\textbf{B}\textbf{B}\textbf{v}/\mu_0)+\textbf{v}(p+B^2/2\mu_0)]-\rho \textbf{v} \textbf{g}&=0, \\ \label{mhd4}
\frac{\partial \bf{B}}{\partial t} + \nabla \times (\textbf{v} \times \textbf{B})&=0,\\ \label{mhd5}
\nabla \cdot \textbf{B} &= 0,
\end{align}
where $\rho$ is the mass density, \textbf{B} is the magnetic field strength, $p$ is the gas pressure, $E$ is the total energy density, and \textbf{g} is the acceleration due to gravity acts along the negative $x-$axis of the simulation setup. Here, the ideal gas law $\displaystyle{p=\frac{\rho k_b T}{\mu m_H}}$ is used in the simulation, and the energy density, $\displaystyle{E = \frac{p}{\gamma -1}+\frac{\rho v^2}{2}+ \frac{B^2}{2\mu_0}}$, where, $T$ is the temperature, $\mu_0$ is the magnetic permeability in vacuum, $\mu=0.6$ is the coronal abundance, $m_H$ is the proton mass, $\gamma=5/3$ is the gas constant for the mono-atomic gas, and $k_b$ is the Boltzmann constant. The solenoidal condition of the magnetic field ($\nabla \cdot \textbf{B}=0$) is satisfied by the Powell's scheme.

The simulation represents an open field magnetic structure. The simulation box is implanted in the lower corona with the spatial dimension of $x \in [0, 50]$ Mm, $y \in [-2.5, 2.5]$ Mm, and $z \in [-2.5, 2.5]$ Mm with a spatial resolution of 0.39 Mm, 0.01 Mm, and 0.01 Mm along the $x$, $y$, and $z$ directions respectively. Furthermore, density inhomogeneities are placed randomly along the $y-z$ according to 

\begin{align}\label{den_inhomo}
  \rho(x,y,z)= \bigg(\rho_0+ \sum_{i=0}^{50} A_i \exp\big(-[(y-y_i)^2-(z-z_i)^2]/2\sigma_i^2\big)\bigg) \ \exp\big(-x/H(y,z)\big),
\end{align}
where, $\rho_0=2 \times 10^{-16}$ g cm$^{-3}$ is the background density, which is the typical density of the lower solar corona ($\sim$ 1.05 solar radius), $A_i$'s are the density amplitudes which are taken randomly from the uniform distribution of $[0,5]\rho_0$, $\sigma_i$'s are the spatial extent of the density inhomogeneities which are drawn randomly from the uniform distribution $[0,250]$ km, and $H(y,z)$ is the scale height which increases with the temperature and has different values in different locations in the $y-z$ plane. The scale height, in equation (\ref{den_inhomo}) is determined by, $\displaystyle{H(y,z)=\frac{k_b T(y,z)}{\mu m_H g}}$, where, $g=274$ m s$^{-2}$ is the acceleration due to gravity at the solar surface. In the initial condition of the simulation, the gas pressure is given by, $p = p_0 \exp(-x/H(y,z))$, where, $p_0=\beta |{\bf B_b}|^2/2$, plasma-$\beta$=0.15, and the background magnetic field, ${\bf B_b}=B_b{\bf \hat{x}}$, with the magnetic field strength, $B_b=5$ G are used in the simulation. Therefore, the temperature, $T$ is constant along the field lines at the initial state of the simulation. However, the evolution of the gas pressure and temperature are governed by the equations (\ref{mhd2}) and (\ref{mhd3}).

After the initial setup of the simulation, the system is allowed to evolve for $\sim 100$ s before implementing any drivers so that the system comes to a pressure equilibrium state. In this stage, all the boundaries are kept open such that any generated MHD wave can leave the simulation domain. Once the system reaches the pressure equilibrium, the bottom boundary is excited by incorporating velocity drivers given by the following equations:

\begin{align}\label{velo_driver_vy}
    v_y(t) &= \sum_{i=1}^{10} U_i \sin(\omega_i t),\\ \label{velo_driver_vz}
    v_z(t) &= \sum_{i=1}^{10} V_i \sin(\omega_i t).
\end{align}
In this stage, the boundaries in the $y-z$ directions are set to be periodic. Here, $\omega_i$'s are the angular frequencies which are chosen from the observed oscillation periods of transverse waves in the off-limb solar plumes which follows a log-normal distribution in the range of 61 s to 2097 s with a mode value of 121 s \citep{2014ApJ...790L...2T}. $U_i,V_i$ are the velocity amplitudes which are chosen randomly from the uniform distributions of $[-U_0,U_0]$, and $[-V_0,V_0]$, where $U_0=V_0=11/\sqrt{2}$ km s$^{-1}$ \citep[see][for details]{2019ApJ...881...95P}. The total duration of the simulation is 1000 s with a 20 s cadence, which gives the 50 snapshots of the simulation. Incorporation of the transverse velocity drivers at the base of the simulation box generates transverse MHD waves that propagate along the vertical direction ($+x$ direction). Propagation of these waves in the density inhomogeneous medium leads to the generation of the turbulence \citep{2017NatSR...714820M, 2019ApJ...881...95P}. The top boundary of the simulation box is kept open so that a negligible amount of the pointing flux reflects back, and there are no counter-propagating waves. We estimate the temporal evolution of the average density at the open boundary at $x=50$ Mm between $t=0$ to 880 s, which has a maximum variation of 4.68\%. This implies that the mass flux through the open boundary is very small, and hence it does not much affect the density distribution. We also estimate the temporal evolution of the average density of the entire simulation domain and obtained that the maximum change of the average density is 0.367\% between $t=0$ to 880 s. The numerical mass diffusion in a simulation code depends on the resolution of the setup. Higher the resolution, lower the numerical mass diffusion effect. The resolution of the simulation presented in this work is taken as 10 km along the $y-z$ direction, which is sufficiently small. We also notice that the diffusive exchange of mass between the overdense and the underdense structures is significantly less when the simulation is run for $\sim 100$ s before it achieves the equilibrium. However, we believe that increasing the resolution of the simulation generates more sub-structures, but does not alter much the results presented in Section \ref{sec:filling factor}, because we are integrating different structures along the LOS.

\begin{figure}[hbt!]
    \centering
    \gridline{\fig{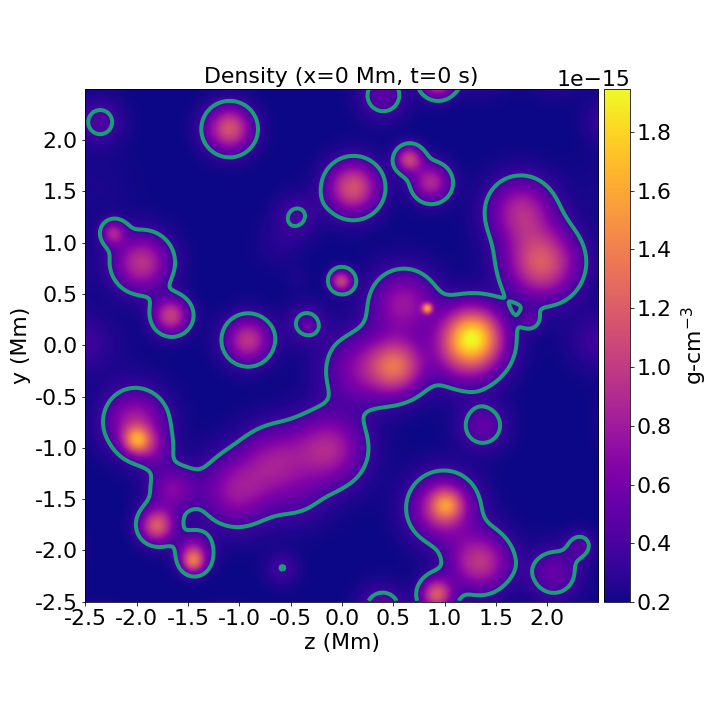}{0.45\textwidth}{(a)}
    {\fig{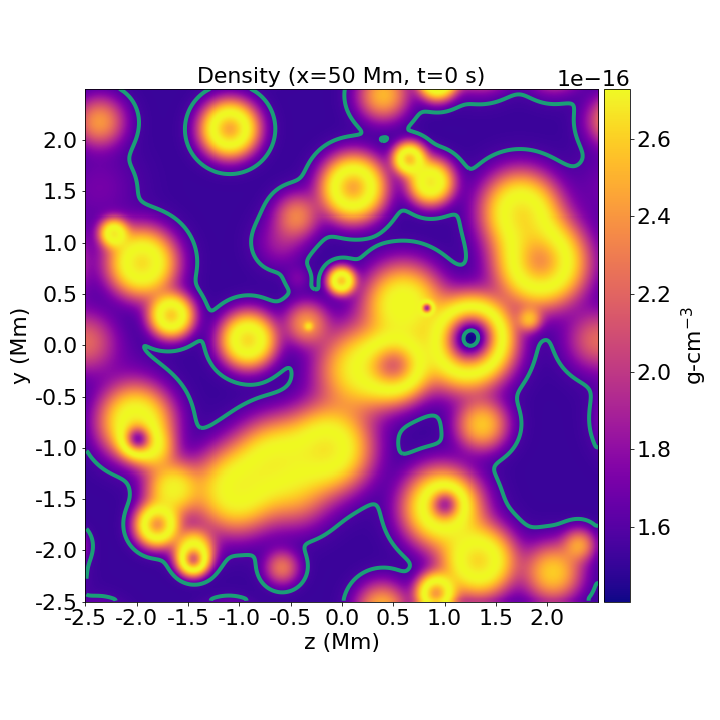}{0.45\textwidth}{(b)}}}
    
    \gridline{\fig{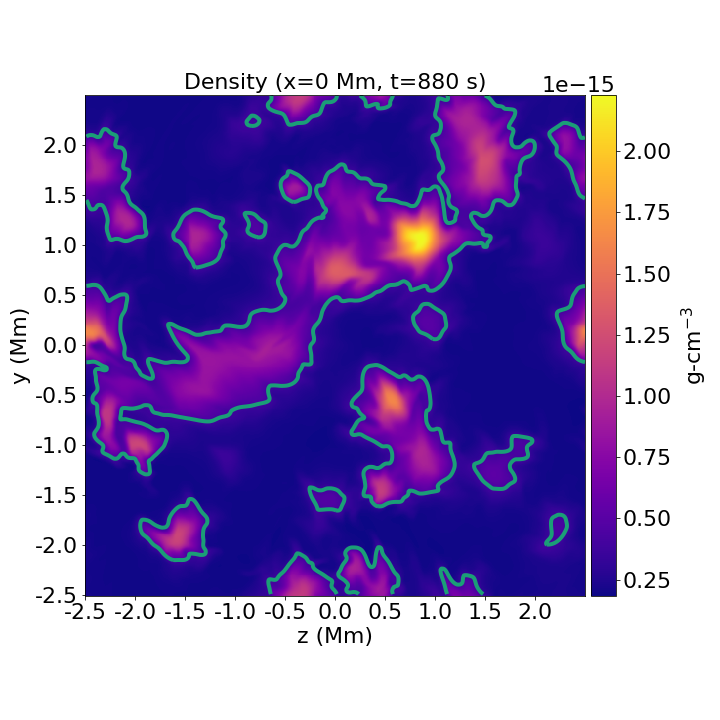}{0.45\textwidth}{(c)}
    {\fig{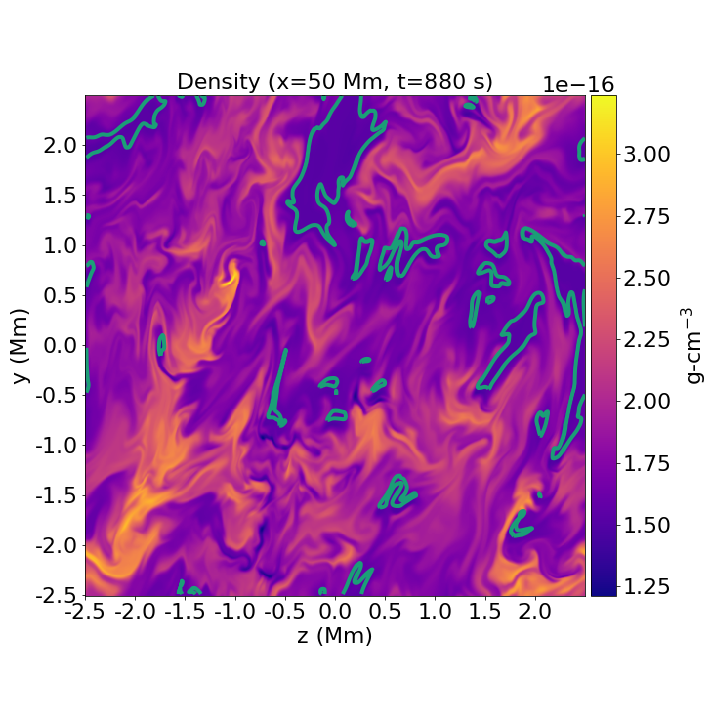}{0.45\textwidth}{(d)}}}
    \caption{\textit{Top panel:} Variation of the density along the $y-z$ plane when the system is in equilibrium ($t=0$), at (a) $x=0$ and (b) $x=50$ Mm obtained from the simulation using MPI-AMRVAC. The contours represent the density threshold, $\rho_{th}(x)$ values for $\lambda_f=0.1$ given by Equation (\ref{eq:density threshold}). \textit{Bottom panel:} Same as the top panel at (c) $x=0$ and (d) $x=50$ Mm when the turbulence is present in the medium ($t=880$ s).}
    \label{fig:den_simu_contour}
\end{figure}

\begin{figure}[hbt!]
    \centering
    \gridline{\fig{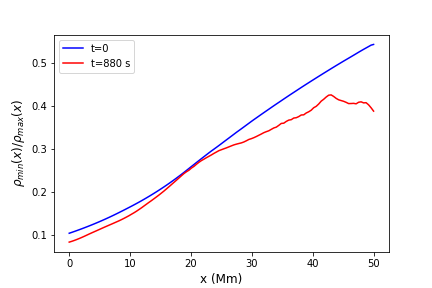}{0.5\textwidth}{(a)}
    {\fig{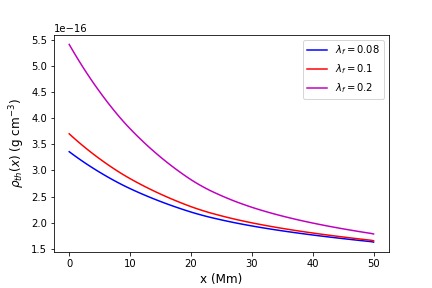}{0.5\textwidth}{(b)}}}
    
    \caption{(a) Variation of the density contrast along the $x-$direction for $t=0$ and 880 s. (b) Variation of density threshold, $\rho_{th}$ at $t=0$ along $x-$direction for different values of $\lambda_f$.}
    \label{fig:den_cont-den_th}
\end{figure}

\begin{figure}
    \centering
    \centering
    \gridline{\fig{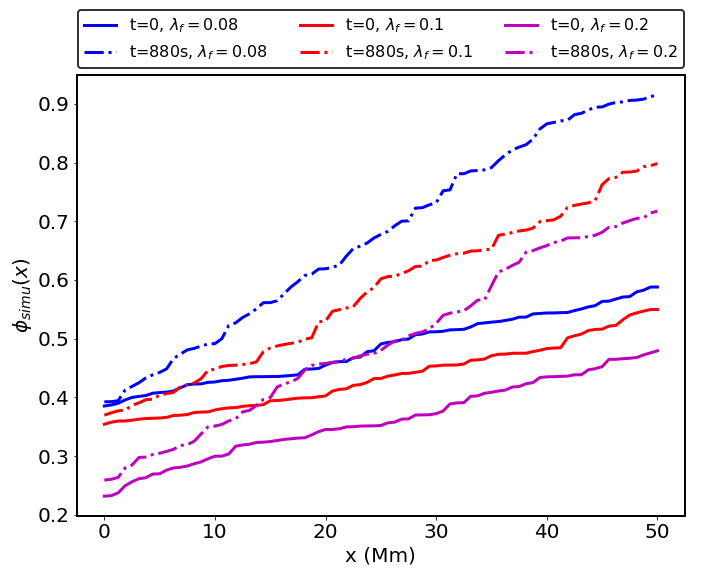}{0.5\textwidth}{(a)}
    {\fig{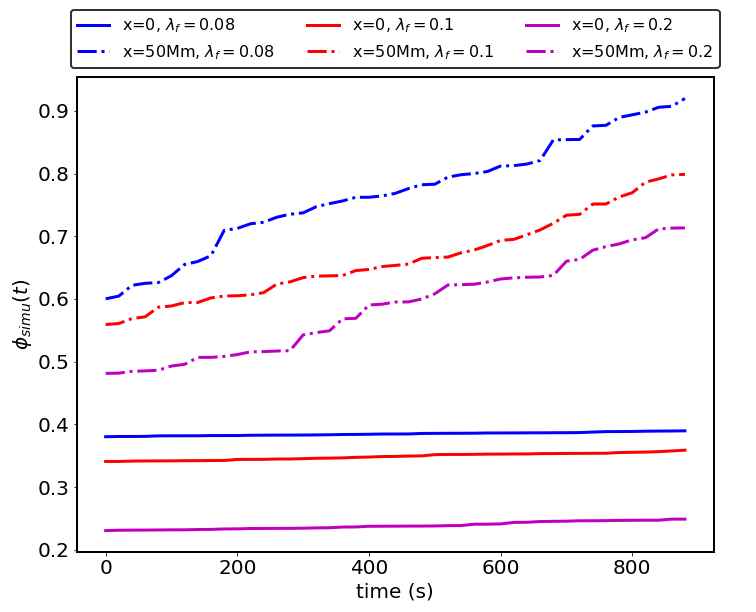}{0.5\textwidth}{(b)}}}
    \caption{(a) Variation of filling factor along $x-$direction for $t=0$ and 880 s for different values of $\lambda_f$. (b) Temporal variation of the density filling for $x=0$ and $x=50$ Mm for different values of $\lambda_f$.}
    \label{fig:simu_ff_compare}
\end{figure}

\begin{figure}[hbt!]
    \centering
    \gridline{\fig{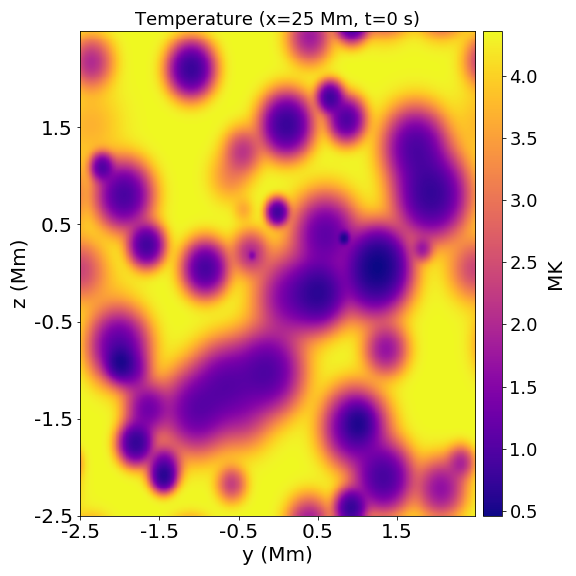}{0.48\textwidth}{(a)}
    {\fig{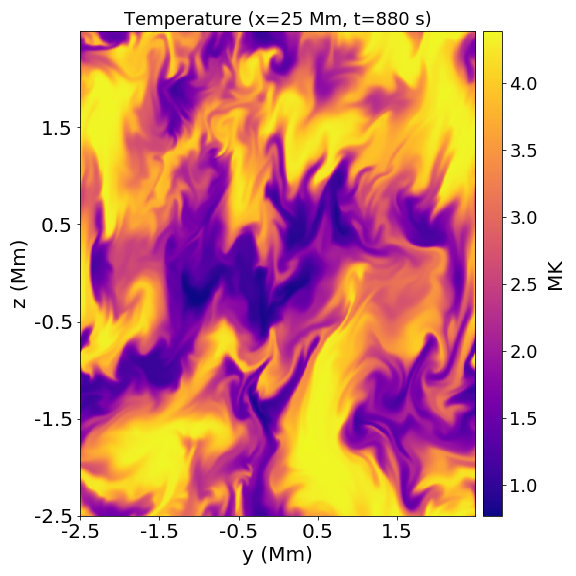}{0.48\textwidth}{(b)}}}
    \caption{Variation of the temperature along the $y-z$ plane at $x=25$ Mm when (a) the system is in equilibrium ($t=0$), and (b) the turbulence is present in the medium ($t=880$ s) obtained from the simulation.}
    \label{fig:temp_simulation}
\end{figure}

\begin{figure}[hbt!]
    \centering
    \gridline{\fig{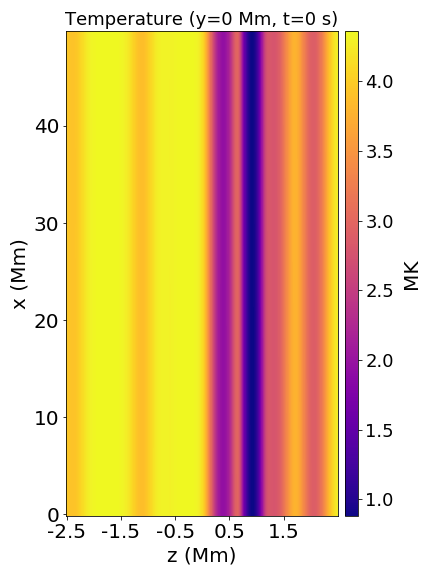}{0.48\textwidth}{(a)}
    {\fig{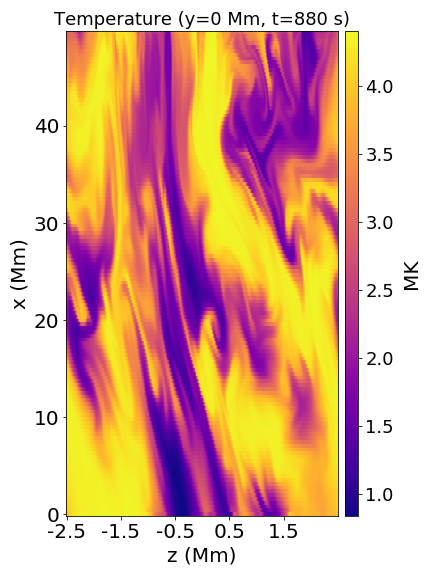}{0.48\textwidth}{(b)}}}
    \caption{Variation of the temperature along the $x-z$ plane at $y=0$ Mm when (a) the system is in equilibrium ($t=0$), and (b) the turbulence is present in the medium ($t=880$ s) obtained from the simulation.}
    \label{fig:temp_simulation_fl}
\end{figure}

\begin{figure}[hbt!]
    \centering
    \includegraphics[width=0.5\columnwidth]{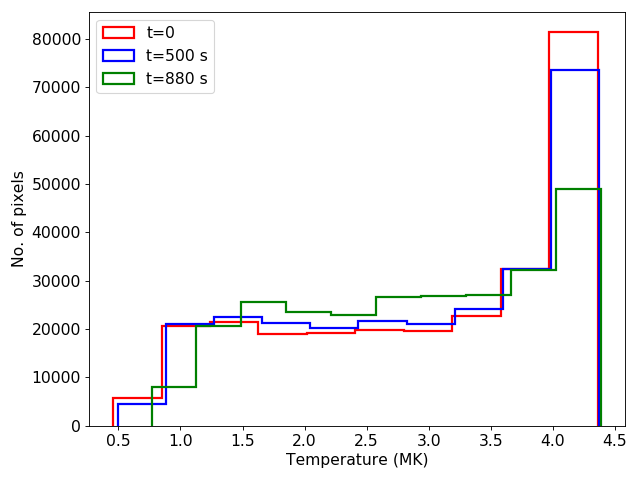}
    \caption{Evolution of the temperature distribution in the $y-z$ plane obtained from the simulation at $x=25$ Mm.}
    \label{fig:temp_hist}
\end{figure}

\begin{figure}[hbt!]
    \centering
    \gridline{\fig{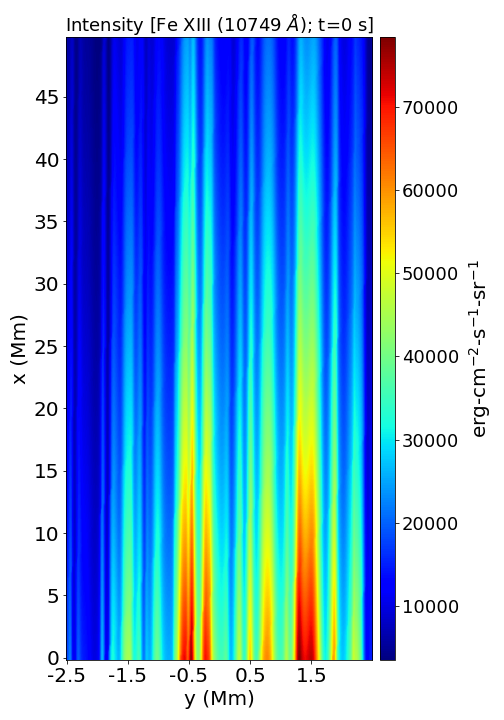}{0.38\textwidth}{(a) LOS=0, $t=0$}\hspace{-3cm}
    {\fig{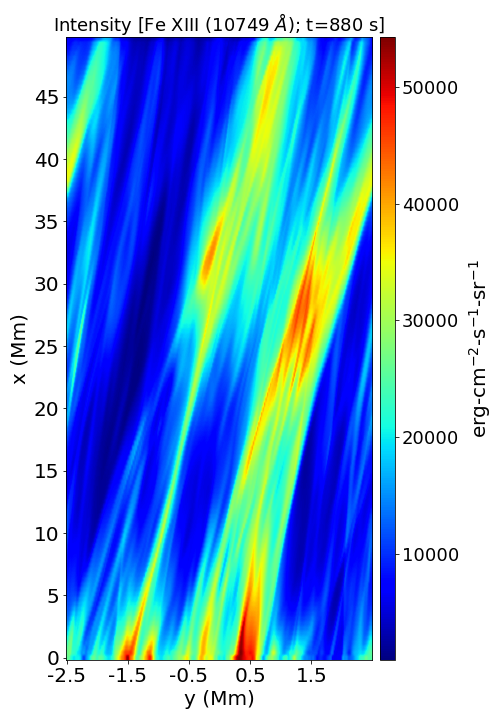}{0.38\textwidth}{(b) LOS=0, $t=880$ s}}}
    
    \gridline{\fig{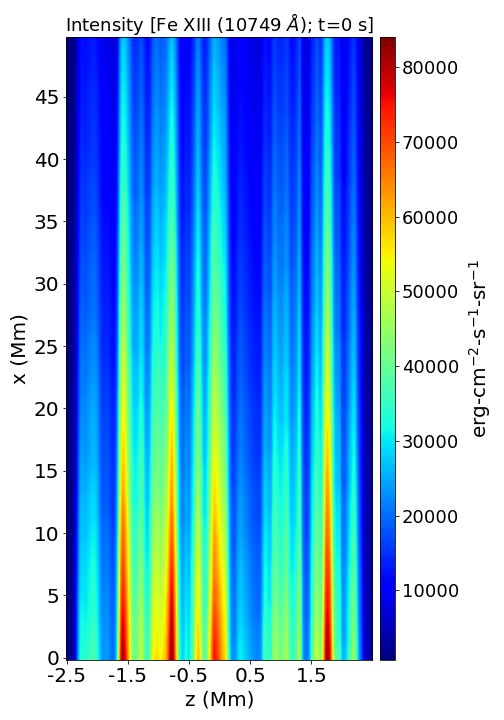}{0.38\textwidth}{(c) LOS=$\pi/2$, $t=0$}\hspace{-3cm}
    {\fig{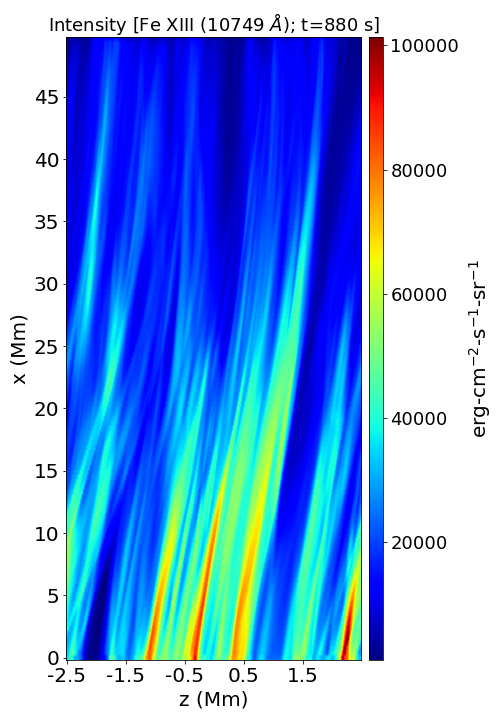}{0.38\textwidth}{(d) LOS=$\pi/2$, $t=880$ s}}}
    \caption{\textit{Top panel:} Intensity map of the synthetic images for LOS=0 obtained by the forward modeling using FoMo for Fe XIII 10749 \AA\ when (a) the system is in the pressure equilibrium ($t=0$), and (b) there is turbulence in the medium ($t=880$ s). \textit{Bottom panel:} Same as top panel with LOS direction $\pi$/2 for (c) $t=0$ and (d) $t=880$ s.}
    \label{fig:synthetic_intensity}
\end{figure}

\begin{figure}[hbt!]
    \centering
    \gridline{\fig{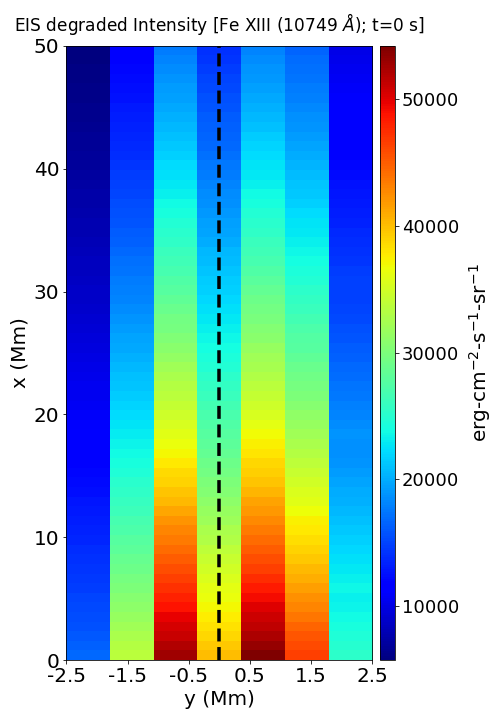}{0.38\textwidth}{(a) LOS=0, $t=0$}\hspace{-2.5cm}
    {\fig{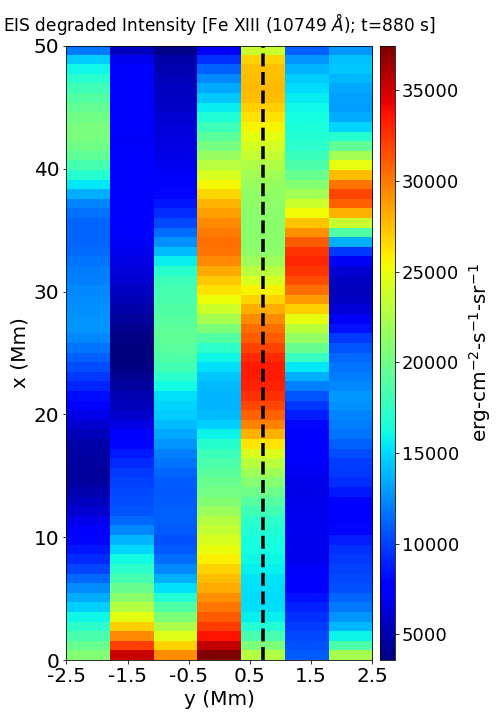}{0.38\textwidth}{(b) LOS=0, $t=880$ s}}}
    
    \gridline{\fig{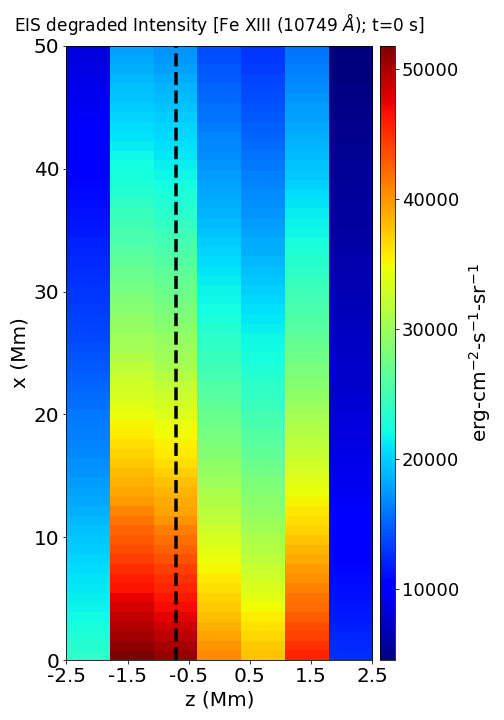}{0.38\textwidth}{(c) LOS=$\pi/2$, $t=0$} \hspace{-2.5cm}
    {\fig{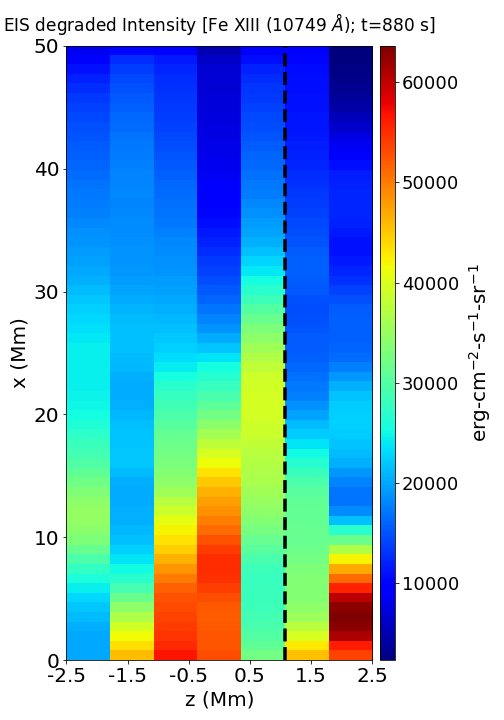}{0.38\textwidth}{(d) LOS=$\pi/2$, $t=880$ s}}}
    \caption{{\textit{Top panel:}} EIS degraded intensity map for LOS=0 for (a) $t=0$ (b) $t=880$ s. {\textit{Bottom panel:}} Same as top panel with LOS=$\pi/2$ for (c) $t=0$ and (d) $t=880$ s. The Dashed black lines represent the cut along the $x$ direction where we estimate the filling factors.}
    \label{fig:eis_synthetic_intensity}
\end{figure}

\begin{figure}[hbt!]
    \centering
    \gridline{\fig{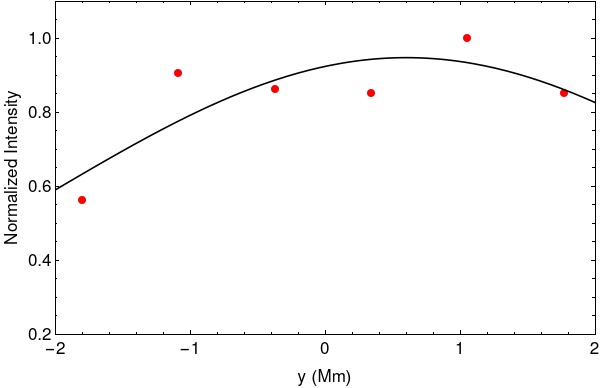}{0.45\textwidth}{(a) LOS=0, $t=0$}
    {\fig{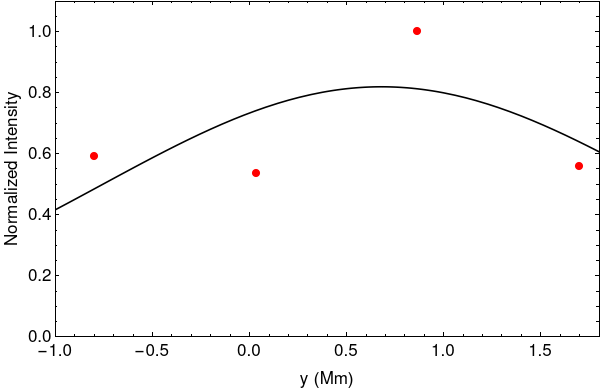}{0.45\textwidth}{(b) LOS=0, $t=880$ s}}
    }
    
    \gridline{\fig{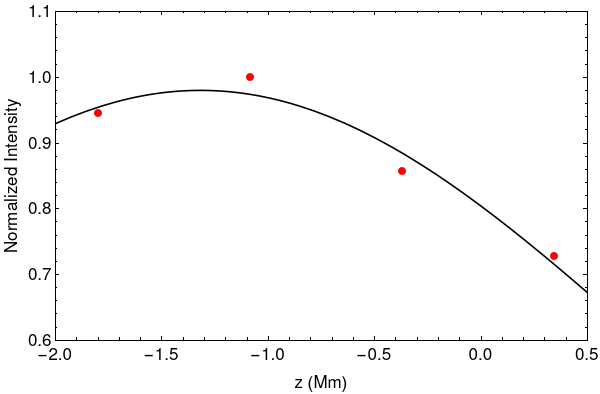}{0.45\textwidth}{(c) LOS=$\pi/2$, $t=0$, }
    {\fig{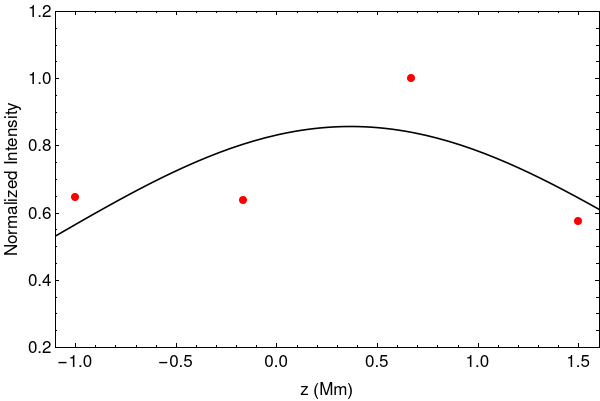}{0.45\textwidth}{(d) LOS=$\pi/2$, $t=880$ s}}
    }
    \caption{Gaussian fit of the intensity across the plasma columns at $x=25$ Mm. The red dots represent the EIS degraded intensity across the plasma columns and the solid black lines are the Gaussian fits of the intensity.}
    \label{fig:gaussian_fit}
\end{figure}

\begin{figure}[hb!]
    \centering
    \includegraphics[scale=0.4]{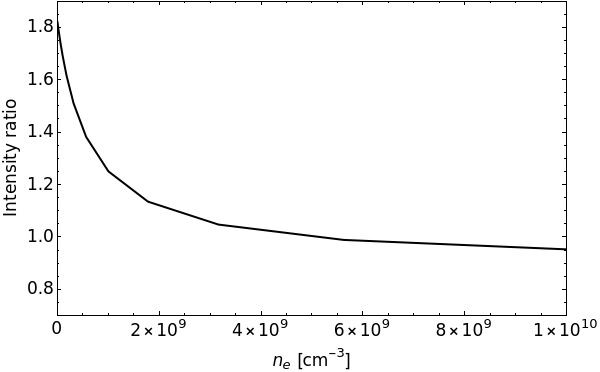}
    \caption{Calibration of the intensity ratio with electron number density, $n_e$ for the density sensitive lines: Fe XIII 10749 \AA\ and 10800 \AA\ obtained from CHIANTI (version 9).}
    \label{fig:chianti_calibration}
\end{figure}

\begin{figure}[hbt!]
    \centering
    \gridline{\fig{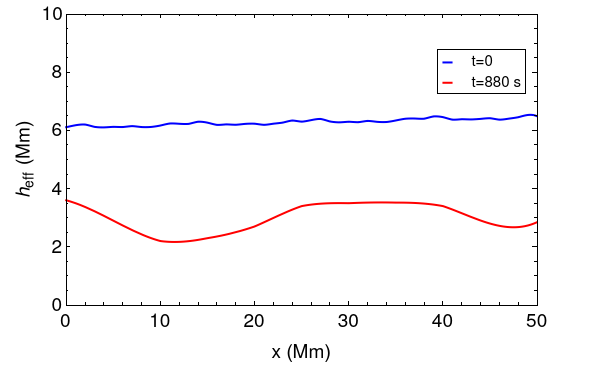}{0.5\textwidth}{(a) LOS=0}
    {\fig{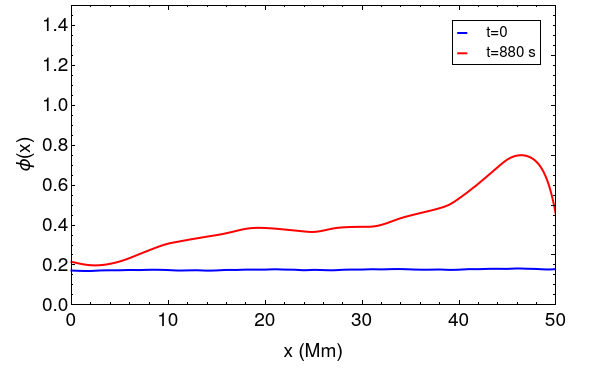}{0.5\textwidth}{(b) LOS=0}}}
    
    \gridline{\fig{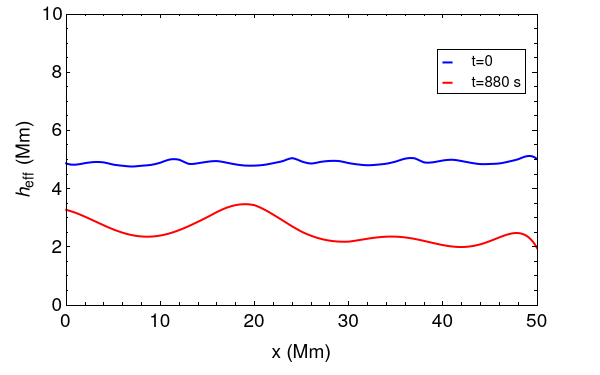}{0.5\textwidth}{(c) LOS=$\pi/2$} 
    {\fig{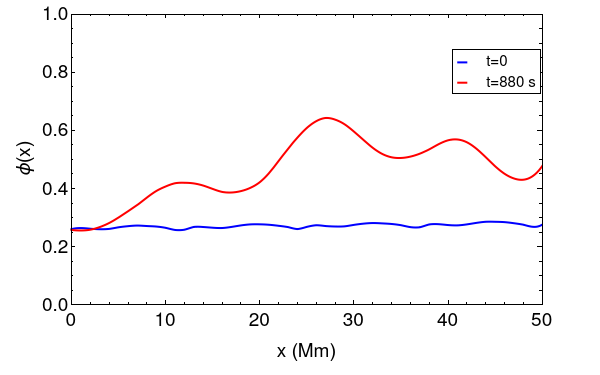}{0.5\textwidth}{(d) LOS=$\pi/2$}}}
    \caption{Variation of $h_{eff}$ and filling factor along $x$ direction obtained from forward modeling using FoMo. (a) and (c) represent the variation of $h_{eff}$, (b) and (d) represent the variation of filling factor along the black dashed line marked in the Figure \ref{fig:eis_synthetic_intensity} for LOS directions 0 and $\pi/2$ respectively.}
    \label{fig:h-phi}
\end{figure}

\section{Method and results}\label{sec:filling factor}

We define the density filling factor as the fraction of the volume occupied by the overdense plasma structures with respect to the total volume of the region. The filling factor gives an estimation of the inhomogeneity of the medium. We have estimated the density filling factor by two different methods described in the following sub-sections.

\subsection{Estimation of filling factor using MHD simulation}\label{subsec:filling factor from area}

We calculate the density filling factor of the overdense regions obtained from the MHD simulation by taking the fraction of the area filled with plasma having a density greater than the chosen density threshold, $\rho_{th}(x)$ to the area of the $y-z$ plane. To find the area of the region which has the density of more than $\rho_{th}(x)$ at each plane along $x$, we define the contours equal to the density threshold $\rho_{th}(x)$ given by

\begin{align}\label{eq:density threshold}
    \rho_{th}(x) = \rho_{min}(x) + \lambda_f \big(\rho_{max}(x)-\rho_{min}(x)\big), 
\end{align}
 where, $\rho_{min}(x)$ and $\rho_{max}(x)$ are the minimum and maximum densities respectively. The density maps obtained from the simulation are shown in Figure \ref{fig:den_simu_contour} at $x=0$ and $x=50$ Mm, for two different times, $t=$0, 880 s. The contours represent the density threshold, $\rho_{th}(x)$ estimated from equation (\ref{eq:density threshold}). At $t=0$, when the system is in pressure equilibrium, there is a variation of $\rho_{min}(x)/\rho_{max}(x)$ along the $x-$direction. This is due to the presence of the density scale height, $H(y,z)$ term in the density inhomogeneity formulation in Equation (\ref{den_inhomo}). The $H(y,z)$ increases with the temperature and has different values at different locations in the $y-z$ plane. The temperature inside the density enhanced regions are less compared to the background medium, and the temperature variation along the field lines is quite small in the pressure equilibrium state. Hence, the scale height of the overdense regions are smaller than background medium. Therefore, the variation of the density along the $x-$direction in the overdense regions are faster than the low density regions, and hence the $\rho_{min}(x)/\rho_{max}(x)$ varies along the $x-$direction at $t=0$. The variation of the $\rho_{min}(x)/\rho_{max}(x)$ along the $x-$direction is shown in Figure \ref{fig:den_cont-den_th}(a) for two different times, when the system is in equilibrium ($t=0$), and when the turbulence is present in the medium ($t=880$ s). $\rho_{th}(x)$ in Equation (\ref{eq:density threshold}) is prescribed in such a way that $\rho_{min}(x)<\rho_{th}(x)<\rho_{max}(x)$, which is valid only if $\displaystyle{0<\lambda_f<1}$. Here, $\lambda_f$ is a free parameter that decides the area of coverage. We choose $\lambda_f=$0.08, 0.1, and 0.2 for estimating the density filling factor of the medium. The variation of $\rho_{th}(x)$ along the $x-$direction is shown in Figure \ref{fig:den_cont-den_th}(b) for different $\lambda_f$. Larger the value of $\lambda_f$, larger the density threshold, $\rho_{th}(x)$ (see Equation \ref{eq:density threshold}). Hence, for a larger $\lambda_f$, the density contours separate out lesser area of the overdense regions from the background and hence the filling factor drops down. Similarly, for smaller values of $\lambda_f$ the density filling factor increases. 
We estimate the value of $\displaystyle{\rho_{th}(x)}$ for $t=0$, and find it to be decreasing with $x$. This is the consequence of density stratification due to gravity. The area of the region whose density is more than $\rho_{th}(x)$ on $y-z$ plane is calculated by setting binary masks of density. For that, we create binary masks in the $y-z$ plane and assign `1' for the pixels that have a density more than $\rho_{th}$, and `0' for the pixels with density less than $\rho_{th}$. Then, we count the number of pixels that are assigned as `1'. This gives the area of the region whose density is more than $\rho_{th}$ in the units of pixel numbers of the simulation grid of $y-z$ plane. Finally, the filling factor, $\phi_{simu}(x)$ in the simulation is given by

\begin{align}\label{eq: filling factor simulation}
   \phi_{simu}(x) = \frac{\text{Area of the region whose density is more than}\ \rho_{th}(x)}{\text{Area of the $y-z$ plane}}.
\end{align}

The variation of the filling factor along the $x$ direction is shown in Figure \ref{fig:simu_ff_compare}(a). Here, we calculate the cross-sectional area due to all the plasma columns which increase with $x$ when the system is in equilibrium ($t=0$), and hence the density filling factor of the medium also increases by the same fraction. For the turbulent medium ($t= $880 s), the density filling factor increases with $x$ more rapidly due to the generation of turbulence which leads to the mixing and distortion of the plasma column boundaries. This lead to a more homogeneous density variation leading to the increase in the density filling factor. The temporal variation of the density filling factor for different heights and $\lambda_f$ are shown in Figure \ref{fig:simu_ff_compare}(b). The variation of the filling factor with time is very small at the bottom boundary of the $x$ plane ($x=0$) where the velocity drivers are incorporated. At $x=0$, the filling factor varies from 0.38 to 0.39 for $\lambda_f=0.08$, 0.34 to 0.36 for $\lambda_f=0.1$, and 0.23 to 0.25 for $\lambda_f=0.2$ with time, $t=0$ to 880 s. Due to the wave-driven turbulence, the filling factor variation is more with time at a higher height. At $x=50$ Mm, the filling factor increases from 0.59 to 0.92 for $\lambda_f=0.08$, 0.55 to 0.80 for $\lambda_f=0.1$, and 0.48 to 0.72 for $\lambda_f=0.2$ with time, $t=0$ to 880 s.

To estimate the temperature of the medium, we use the pressure, $p$, the background density, $\rho_0$, and the density contrast values (from equation \ref{den_inhomo}) using the ideal gas law. This gives the initial temperature of the medium in the range of $\approx 0.4-4$ MK. An example of the variation of the temperature distribution at $x=25$ Mm along the $y-z$ plane are shown in Figure \ref{fig:temp_simulation} for $t=0$ and 880 s. This represents that the regions with higher densities consist of lower temperatures, and the regions with lower densities consist of higher temperatures. Figure \ref{fig:temp_simulation_fl} shows the temperature along the field lines ($x-z$ plane) at $y=0$ for $t=0$ and 880 s, which shows that the variation of the temperature along the filed lines are very small in the pressure equilibrium state ($t=0$). Furthermore, when the turbulence is generated in the medium due to the wave excitation, the overdense regions mix up with the background, and hence the temperature distribution becomes more homogeneous in space (see Figures \ref{fig:temp_simulation}(b) and \ref{fig:temp_simulation_fl}(b)). Figure \ref{fig:temp_hist} shows the comparison of the temperature distribution in the $y-z$ plane at different times at $x=25$ Mm. This reflects the fact that the temperature distribution for $t=880$ s is more homogeneous than $t=500$ and 0 s. We have estimated that the minimum and maximum temperatures of the medium at $t=0$ are 0.452 MK and 4.36 MK, which increases to 0.812 MK and 4.41 MK at $t=880$ s respectively. One of the possibilities of this small increase of the temperature is due to the adiabatic compression and rarefaction of the medium due to the wave-driven turbulence. However, heat conduction, and optically thin radiative losses terms are not incorporated in the simulation set up in the Equations (\ref{mhd1}-\ref{mhd5}). Certainly, these terms play roles in establishing the coronal density structures, but, the main aim of this work is to investigate the effect of the wave-driven turbulence only in the density inhomogeneities of the solar corona without accounting the conduction and radiative loss terms. We understand that this is the limitation of this study, but nevertheless, it is interesting because it shows that MHD turbulence alone is able to increase filling factors in the solar corona. A more detailed investigation incorporating the energy losses and transport can be incorporated.

\subsection{Estimation of filling factor from synthetic spectra using FoMo} \label{subsec:filling factor-fomo}

To convert the physical variables obtained from the simulation (e.g. density, velocity, energy density) into spectroscopic observables, e.g. specific intensity, we use forward modeling using FoMo \citep{2013A&A...555A..74A, 2016FrASS...3....4V}. FoMo calculates the line of sight (LOS) integrated emission from the optically thin coronal plasma. Here we define the LOS$=0$, and $\pi/2$ are along the $+z$ and  $-y$ directions respectively. In this method, we generate synthetic spectra for the density sensitive lines pairs, Fe XIII 10749 \AA\ and 10800 \AA\ \citep{2016JGRA..121.8237L} for two LOS directions: $0$ and $\pi/2$ when the system is in the pressure equilibrium ($t=0$), and when the turbulence is present in the medium ($t=880$ s) (see Figure \ref{fig:synthetic_intensity}). Next, we degrade the generated synthetic images into the spatial resolution of 720 km along the $x$ and $y$ directions, which is comparable with the spatial resolution ($1^"$/pixel) of EIS \citep{2007SoPh..243...19C}. The EIS degraded synthetic images are shown in Figure \ref{fig:eis_synthetic_intensity} for two different LOS directions and $t$. Following \cite{2015ApJ...800..140G}, the emission measure, EM, can be calculated as, 
\begin{align}
   \label{eq:EM1}
    EM_1 &= \frac{I_{EIS}}{0.83 A_b G(n_e,T)}. 
\end{align}
Here, $G(n_e,T)$ is the contribution function which we calculate in a 3D grid of $(n_e,T)$ within the domain, $n_e=10^8-10^{11}$ cm$^{-3}$, and log $T=5.5-6.95$ K for the density sensitive lines: Fe XIII 10749\AA\ and 10800\AA. For calculating $G(n_e,T)$, we use {\it goft\_table.pro} from {\it FoMo-idl} \citep{2016FrASS...3....4V}, and {\sf CHIANTI} version 9 \citep{1997A&AS..125..149D, 2019ApJS..241...22D}. $I_{EIS}$ is the EIS degraded synthetic intensity for Fe XIII 10749 \AA\ line obtained from FoMo, the temperature, log($T$)=6.25 K is taken as the peak formation temperature for the Fe XIII 10749 line, where the contribution function has the maximum value, and $A_b=0.6$ is the coronal abundance. Using the intensity ratio of the EIS degraded synthetic maps of Fe XIII 10749 \AA\ and 10800 \AA\ lines, and the intensity vs density calibration obtained from CHIANTI (see Figure \ref{fig:chianti_calibration}), we calculate the electron density, $n_e$ in the projected plane perpendicular to the LOS direction. The emission measure can be estimated by
 
 \begin{align}\label{eq:EM2}
    EM_2 &= \int_{h_{eff}} n_e^2 {\rm d}h,
\end{align}
where, $h_{eff}$ is the LOS depth of the overdense plasma regions of the EIS degraded synthetic maps. In this case, the $h_{eff}$ at each height, $x$ is measured assuming the overdense plasma region has an azimuthal symmetry around the $x$ direction. Hence, the effective depth of the overdense plasma regions, $h_{eff}$ is calculated by fitting Gaussian with the intensity across the overdense plasma regions at different heights and taking the FWHM of the Gaussian fit. Thus, the plasma filling factor in the plane perpendicular to the LOS direction can be calculated by \citep{ 2015ApJ...800..140G}
\begin{align}\label{eq:filling factor}
    \phi = \frac{I_{EIS}}{0.83 A_b G(n_e,T) n_e^2\ h_{eff}}.
\end{align}    

 The intensity and the Gaussian fit across the plasma columns at $x=25$ Mm are shown in Figure \ref{fig:gaussian_fit} for two different times and LOS directions.  The variation of $h_{eff}$ along the $x$ direction are shown in the Figures \ref{fig:h-phi}(a) and (c) for LOS directions 0 and $\pi/2$ respectively. For calculating the $h_{eff}$, the FWHM of the intensity Gaussian fit mainly depends on two factors: (i) the extension of the Gaussian distribution along the $y$ direction, and (ii) how much the Gaussian distribution is skewed due to the intensity enhancement in the synthetic spectra. E.g. the horizontal spread (along the $y$ direction) of the Gaussian fit  for the plasma column in Figure \ref{fig:eis_synthetic_intensity}(b) is more at $x=0$ ($-2.5$ Mm $\leq$ y $\leq$ 1 Mm) than $x=10$ Mm ($-1.8$ Mm $\leq$ y $\leq$ 1 Mm), and hence the FWHM of the Gaussian fit or $h_{eff}$ drops from 3.2 Mm at $x=0$ to 2 Mm at $x=10$ Mm, which is reflected in  Figure \ref{fig:h-phi}(a) for $t=880$ s. On the other hand, the intensity across the black dashed line in the synthetic map of Figure \ref{fig:eis_synthetic_intensity}(d) is more skewed at $x=25$ Mm than $x=20$ Mm (where $-1$ Mm $\leq$ $z$ $\leq 1.5$ Mm for both cases). Therefore, the Gaussian fit becomes more skewed at $x=25$ Mm, and hence the $h_{eff}$ decreases from 3.4 Mm at $x=20$ Mm to 2.5 Mm at $x=25$ Mm, which reflects in Figure \ref{fig:h-phi}(c). The variation of the filling factors obtained from the EIS degraded synthetic spectra are shown in Figures \ref{fig:h-phi}(b) and (d) for the LOS directions 0 and $\pi/2$ respectively (along the black dashed line marked in Figure \ref{fig:eis_synthetic_intensity}(a)-(d)). We find that the maximum filling factor value is $\approx 0.8$ for the turbulent medium (for LOS=0), whereas, for the equilibrium case, it reduces to $\approx 0.2$. This depicts that the medium becomes more density homogeneous due to the development of turbulence. From Figures \ref{fig:h-phi}(b) and (d), we notice that the filling factor values are different for two LOS directions, and this is more significant when the turbulence is present in the medium. This infers that the density inhomogeneities are irregular and asymmetric around the $x$ direction.

\section{Discussions and conclusion} \label{sec: discussion}

While estimating the density filling factor by taking the ratio of the area of the density enhanced region to the area of the $y-z$ plane (method described in \ref{subsec:filling factor from area}), we notice that the density filling factor increases with height for both $t=0$ and 880 s, which are shown in Figure \ref{fig:h-phi}(a). However, the increase in the filling factor is larger for the latter case. In the pressure equilibrium state ($t=0$), the overdense plasma columns expand with height in the corona because the total pressure scale height inside the plasma columns which is due to the magnetic and plasma pressure is more than the pressure scale height of the background medium, and hence the background pressure drops more rapidly along the height than the pressure inside the overdense plasma columns. This leads to the expansion of the overdense plasma columns along the transverse direction, and therefore the density filling factor of the medium increases with height. For $t=0$, we estimate that the total cross-sectional area due to all the overdense plasma columns, and hence the density filling factor increases along $x$ direction (see subsection \ref{subsec:filling factor from area} for the estimated values). When the turbulence is present in the medium ($t=880$ s), the plasma columns containing over dense regions disrupt and the density enhanced regions start to mix up with the background medium causing a further rise of the density filling factor. This makes the medium more density homogeneous. For $t=880$ s, the area of the over-dense regions, and hence the density filling factor increases along $x$ direction (see subsection \ref{subsec:filling factor from area} for the estimated values). The variation of the density filling factor with height along coronal loops has been reported by \cite{2009ApJ...694.1256T}. They found the filling factor varies from $0.02$ at the loop footpoints to $0.80$ at $40$ Mm height using the Fe XII spectral lines. For estimating the filling factor by forward modeling, the synthetic spectra which we generate are the LOS integrated intensity obtained from FoMo. The plasma column regions with enhanced densities are located randomly along the $y-z$ plane and extended along the $x$ direction are the major sources of the plasma emission. We calculate the thickness of these plasma emitting regions ($h_{eff}$) by taking the FWHM of the intensity Gaussian fit across the plasma columns. When the cylindrical overdense plasma column structures distort due to turbulence, it is possible that some parts of the plasma emitting regions squeeze and some other parts expand along the transverse direction, and hence $h_{eff}$ decreases or increases accordingly. Therefore, the variation of $h_{eff}$ along the $x$ direction is more significant for the turbulent medium than the equilibrium states which are reflected in Figures \ref{fig:h-phi}(b) and (d). The presence of turbulence might be responsible for the generation of smaller substructures in the medium, or different substructures can mix up with the background medium. This might also be a reason for the large variation of the filling factor along the $x$ direction for a turbulent medium. LOS direction is also a crucial parameter for estimating the density filling factor from forward modeling using FoMo. Due to the presence of turbulence, the plasma columns become irregular in shape and non-axisymmetric around the axis. Therefore, we find that the values of the density filling factors along the black dashed lines in Figures \ref{fig:eis_synthetic_intensity}(b) and (d) are different for LOS=0 and $\pi/2$ (Figures \ref{fig:h-phi}(b) and (d)). However, we find that the filling factor along $x$ direction for the turbulent medium ($t=880$ s) is more than the equilibrium state ($t=0$). It infers that the medium becomes more density homogeneous due to the presence of turbulence. The estimated filling factors we obtain from both methods are in good agreement with previous studies. E.g. \cite{2016ApJ...829...42H} have estimated the filling factor for quiet-Sun and interplume regions of coronal holes using ``two-density slab model", where they have estimated the filling factors are $\approx 10\%-20 \%$. From the spectroscopic observation, the filling factor is found to be $0.10$ for an active region loop \citep{2008ApJ...686L.131W}, and $0.30$ for a cooling loop in quiescent active regions \citep{2009ApJ...695..221L}. For a fan loop structure, \cite{2012ApJ...744...14Y} found the filling factor to be $0.03$ to $0.30$. The filling factor along an entire coronal loop is estimated by \cite{2015ApJ...800..140G} using the spectroscopic observation, where they found the filling factor is increasing along the loop length from $\approx 0.11$ at the footpoint of the loop and exceeds beyond 1 at the loop top. The authors claim that the estimation of filling factors using spectroscopic observations depends on the good background subtraction and the LOS depth of the emitting plasma, which might be the cause of the errors in the above measurements at higher heights. In our study, the density inhomogeneities are incorporated as an initial condition in the simulation setup which follows Gaussian distributions along the $y-z$ plane (according to Equation \ref{den_inhomo}). Therefore, due to the emission from individual overdense plasma columns, the intensity distribution along the $y-z$ plane follow the Gaussian distribution when the system is in equilibrium. But when the turbulence develops in the system, the overdense plasma columns disrupt and hence the intensity distribution does not remain like Gaussian. On the other hand, due to the LOS superposition of the overdense plasma columns, the intensity pattern across the plasma columns does not exactly follow the Gaussian distribution. However, even if there is no LOS superposition of the plasma columns, the intensity distribution across the plasma columns still may not be Gaussian. These effects might cause the error in fitting and hence in calculating $h_{eff}$.
In this study, we have calculated the filling factor by degrading the synthetic images into the resolution of EIS (1$^"$/pixel). We can also extend this work by degrading the synthetic spectra into higher resolution images, e.g. the ``{\it Daniel K. Inouye Solar Telescope}" \citep[DKIST;][]{2021SoPh..296...70R}, which has an angular resolution of $0.1^"$. These higher resolution synthetic maps will provide the information of the finer structures and will lead to investigate the density inhomogeneities of the structures with smaller spatial scales of the medium. But, these are out of the scope of this paper, and we plan to study them in the future.

The study of the density filling factor in the solar corona helps us to understand the overall density structure of the solar corona, and how different is the density structure from the homogeneous approximation. The effect of the turbulence on the density filling factor is presented in our work that can provide scope to compare with the solar coronal structures with the further advancement of the observations. \cite{2014ApJ...795...18V} provides an analytical relation in terms of filling factor that gives the correction term of the energy flux of Alfv{\'e}n waves. Hence, the estimation of the filling factor is crucial for estimating the energy flux in coronal waves which plays an important role in coronal heating.







\bibliography{paper_cor_wave}{}

\begin{thebibliography}{}
\expandafter\ifx\csname natexlab\endcsname\relax\def\natexlab#1{#1}\fi
\providecommand{\url}[1]{\href{#1}{#1}}
\providecommand{\dodoi}[1]{doi:~\href{http://doi.org/#1}{\nolinkurl{#1}}}
\providecommand{\doeprint}[1]{\href{http://ascl.net/#1}{\nolinkurl{http://ascl.net/#1}}}
\providecommand{\doarXiv}[1]{\href{https://arxiv.org/abs/#1}{\nolinkurl{https://arxiv.org/abs/#1}}}

\bibitem[{{Antolin} \& {Van Doorsselaere}(2013)}]{2013A&A...555A..74A}
{Antolin}, P., \& {Van Doorsselaere}, T. 2013, \aap, 555, A74,
  \dodoi{10.1051/0004-6361/201220784}

\bibitem[{{Arregui}(2015)}]{2015RSPTA.37340261A}
{Arregui}, I. 2015, Philosophical Transactions of the Royal Society of London
  Series A, 373, 20140261, \dodoi{10.1098/rsta.2014.0261}

\bibitem[{{Banerjee} {et~al.}(2020){Banerjee}, {Krishna Prasad}, {Pant},
  {McLaughlin}, {Antolin}, {Magyar}, {Ofman}, {Tian}, {Van Doorsselaere}, {De
  Moortel}, \& {Wang}}]{2020arXiv201208802B}
{Banerjee}, D., {Krishna Prasad}, S., {Pant}, V., {et~al.} 2020, arXiv
  e-prints, arXiv:2012.08802.
\newblock \doarXiv{2012.08802}

\bibitem[{{Cranmer} \& {van Ballegooijen}(2005)}]{2005ApJS..156..265C}
{Cranmer}, S.~R., \& {van Ballegooijen}, A.~A. 2005, \apjs, 156, 265,
  \dodoi{10.1086/426507}

\bibitem[{{Culhane} {et~al.}(2007){Culhane}, {Harra}, {James}, {Al-Janabi},
  {Bradley}, {Chaudry}, {Rees}, {Tandy}, {Thomas}, {Whillock}, {Winter},
  {Doschek}, {Korendyke}, {Brown}, {Myers}, {Mariska}, {Seely}, {Lang}, {Kent},
  {Shaughnessy}, {Young}, {Simnett}, {Castelli}, {Mahmoud}, {Mapson-Menard},
  {Probyn}, {Thomas}, {Davila}, {Dere}, {Windt}, {Shea}, {Hagood}, {Moye},
  {Hara}, {Watanabe}, {Matsuzaki}, {Kosugi}, {Hansteen}, \&
  {Wikstol}}]{2007SoPh..243...19C}
{Culhane}, J.~L., {Harra}, L.~K., {James}, A.~M., {et~al.} 2007, \solphys, 243,
  19, \dodoi{10.1007/s01007-007-0293-1}

\bibitem[{{De Moortel} \& {Pascoe}(2012)}]{2012ApJ...746...31D}
{De Moortel}, I., \& {Pascoe}, D.~J. 2012, \apj, 746, 31,
  \dodoi{10.1088/0004-637X/746/1/31}

\bibitem[{{Dere}(2008)}]{2008A&A...491..561D}
{Dere}, K.~P. 2008, \aap, 491, 561, \dodoi{10.1051/0004-6361:200810000}

\bibitem[{{Dere} {et~al.}(2019){Dere}, {Del Zanna}, {Young}, {Landi}, \&
  {Sutherland}}]{2019ApJS..241...22D}
{Dere}, K.~P., {Del Zanna}, G., {Young}, P.~R., {Landi}, E., \& {Sutherland},
  R.~S. 2019, \apjs, 241, 22, \dodoi{10.3847/1538-4365/ab05cf}

\bibitem[{{Dere} {et~al.}(1997){Dere}, {Landi}, {Mason}, {Monsignori Fossi}, \&
  {Young}}]{1997A&AS..125..149D}
{Dere}, K.~P., {Landi}, E., {Mason}, H.~E., {Monsignori Fossi}, B.~C., \&
  {Young}, P.~R. 1997, \aaps, 125, 149, \dodoi{10.1051/aas:1997368}

\bibitem[{{Dmitruk} {et~al.}(2001){Dmitruk}, {Matthaeus}, {Milano}, \&
  {Oughton}}]{2001PhPl....8.2377D}
{Dmitruk}, P., {Matthaeus}, W.~H., {Milano}, L.~J., \& {Oughton}, S. 2001,
  Physics of Plasmas, 8, 2377, \dodoi{10.1063/1.1344563}

\bibitem[{{Einaudi} {et~al.}(1996){Einaudi}, {Velli}, {Politano}, \&
  {Pouquet}}]{1996ApJ...457L.113E}
{Einaudi}, G., {Velli}, M., {Politano}, H., \& {Pouquet}, A. 1996, \apjl, 457,
  L113, \dodoi{10.1086/309893}

\bibitem[{{Goossens} {et~al.}(2012){Goossens}, {Andries}, {Soler}, {Van
  Doorsselaere}, {Arregui}, \& {Terradas}}]{2012ApJ...753..111G}
{Goossens}, M., {Andries}, J., {Soler}, R., {et~al.} 2012, \apj, 753, 111,
  \dodoi{10.1088/0004-637X/753/2/111}

\bibitem[{{Gupta} {et~al.}(2015){Gupta}, {Tripathi}, \&
  {Mason}}]{2015ApJ...800..140G}
{Gupta}, G.~R., {Tripathi}, D., \& {Mason}, H.~E. 2015, \apj, 800, 140,
  \dodoi{10.1088/0004-637X/800/2/140}

\bibitem[{{Hahn} \& {Savin}(2016)}]{2016ApJ...829...42H}
{Hahn}, M., \& {Savin}, D.~W. 2016, \apj, 829, 42,
  \dodoi{10.3847/0004-637X/829/1/42}

\bibitem[{{Klimchuk}(2006)}]{2006SoPh..234...41K}
{Klimchuk}, J.~A. 2006, \solphys, 234, 41, \dodoi{10.1007/s11207-006-0055-z}

\bibitem[{{Landi} {et~al.}(2016){Landi}, {Habbal}, \&
  {Tomczyk}}]{2016JGRA..121.8237L}
{Landi}, E., {Habbal}, S.~R., \& {Tomczyk}, S. 2016, Journal of Geophysical
  Research (Space Physics), 121, 8237, \dodoi{10.1002/2016JA022598}

\bibitem[{{Landi} {et~al.}(2009){Landi}, {Miralles}, {Curdt}, \&
  {Hara}}]{2009ApJ...695..221L}
{Landi}, E., {Miralles}, M.~P., {Curdt}, W., \& {Hara}, H. 2009, \apj, 695,
  221, \dodoi{10.1088/0004-637X/695/1/221}

\bibitem[{{Lemen} {et~al.}(2012){Lemen}, {Title}, {Akin}, {Boerner}, {Chou},
  {Drake}, {Duncan}, {Edwards}, {Friedlaender}, {Heyman}, {Hurlburt}, {Katz},
  {Kushner}, {Levay}, {Lindgren}, {Mathur}, {McFeaters}, {Mitchell}, {Rehse},
  {Schrijver}, {Springer}, {Stern}, {Tarbell}, {Wuelser}, {Wolfson}, {Yanari},
  {Bookbinder}, {Cheimets}, {Caldwell}, {Deluca}, {Gates}, {Golub}, {Park},
  {Podgorski}, {Bush}, {Scherrer}, {Gummin}, {Smith}, {Auker}, {Jerram},
  {Pool}, {Soufli}, {Windt}, {Beardsley}, {Clapp}, {Lang}, \&
  {Waltham}}]{2012SoPh..275...17L}
{Lemen}, J.~R., {Title}, A.~M., {Akin}, D.~J., {et~al.} 2012, \solphys, 275,
  17, \dodoi{10.1007/s11207-011-9776-8}

\bibitem[{{Magyar} {et~al.}(2017){Magyar}, {Van Doorsselaere}, \&
  {Goossens}}]{2017NatSR...714820M}
{Magyar}, N., {Van Doorsselaere}, T., \& {Goossens}, M. 2017, Scientific
  Reports, 7, 14820, \dodoi{10.1038/s41598-017-13660-1}

\bibitem[{{Magyar} {et~al.}(2019){Magyar}, {Van Doorsselaere}, \&
  {Goossens}}]{2019ApJ...882...50M}
---. 2019, \apj, 882, 50, \dodoi{10.3847/1538-4357/ab357c}

\bibitem[{{Matthaeus} {et~al.}(1999){Matthaeus}, {Zank}, {Oughton}, {Mullan},
  \& {Dmitruk}}]{1999ApJ...523L..93M}
{Matthaeus}, W.~H., {Zank}, G.~P., {Oughton}, S., {Mullan}, D.~J., \&
  {Dmitruk}, P. 1999, \apjl, 523, L93, \dodoi{10.1086/312259}

\bibitem[{{McIntosh} \& {De Pontieu}(2012)}]{2012ApJ...761..138M}
{McIntosh}, S.~W., \& {De Pontieu}, B. 2012, \apj, 761, 138,
  \dodoi{10.1088/0004-637X/761/2/138}

\bibitem[{{Morton} {et~al.}(2015){Morton}, {Tomczyk}, \&
  {Pinto}}]{2015NatCo...6.7813M}
{Morton}, R.~J., {Tomczyk}, S., \& {Pinto}, R. 2015, Nature Communications, 6,
  7813, \dodoi{10.1038/ncomms8813}

\bibitem[{{Pant} {et~al.}(2019){Pant}, {Magyar}, {Van Doorsselaere}, \&
  {Morton}}]{2019ApJ...881...95P}
{Pant}, V., {Magyar}, N., {Van Doorsselaere}, T., \& {Morton}, R.~J. 2019,
  \apj, 881, 95, \dodoi{10.3847/1538-4357/ab2da3}

\bibitem[{{Pant} \& {Van Doorsselaere}(2020)}]{2020ApJ...899....1P}
{Pant}, V., \& {Van Doorsselaere}, T. 2020, \apj, 899, 1,
  \dodoi{10.3847/1538-4357/aba429}

\bibitem[{{Parker}(1988)}]{1988ApJ...330..474P}
{Parker}, E.~N. 1988, \apj, 330, 474, \dodoi{10.1086/166485}

\bibitem[{{Peter} {et~al.}(2005){Peter}, {Gudiksen}, \&
  {Nordlund}}]{2005ESASP.596E..14P}
{Peter}, H., {Gudiksen}, B.~V., \& {Nordlund}, A. 2005, in ESA Special
  Publication, Vol. 596, Chromospheric and Coronal Magnetic Fields, ed. D.~E.
  {Innes}, A.~{Lagg}, \& S.~A. {Solanki}, 14.1

\bibitem[{{Porth} {et~al.}(2014){Porth}, {Xia}, {Hendrix}, {Moschou}, \&
  {Keppens}}]{2014ApJS..214....4P}
{Porth}, O., {Xia}, C., {Hendrix}, T., {Moschou}, S.~P., \& {Keppens}, R. 2014,
  \apjs, 214, 4, \dodoi{10.1088/0067-0049/214/1/4}

\bibitem[{{Rappazzo} {et~al.}(2008){Rappazzo}, {Velli}, {Einaudi}, \&
  {Dahlburg}}]{2008ApJ...677.1348R}
{Rappazzo}, A.~F., {Velli}, M., {Einaudi}, G., \& {Dahlburg}, R.~B. 2008, \apj,
  677, 1348, \dodoi{10.1086/528786}

\bibitem[{{Rast} {et~al.}(2021){Rast}, {Bello Gonz{\'a}lez}, {Bellot Rubio},
  {Cao}, {Cauzzi}, {Deluca}, {de Pontieu}, {Fletcher}, {Gibson}, {Judge},
  {Katsukawa}, {Kazachenko}, {Khomenko}, {Landi}, {Mart{\'\i}nez Pillet},
  {Petrie}, {Qiu}, {Rachmeler}, {Rempel}, {Schmidt}, {Scullion}, {Sun},
  {Welsch}, {Andretta}, {Antolin}, {Ayres}, {Balasubramaniam}, {Ballai},
  {Berger}, {Bradshaw}, {Campbell}, {Carlsson}, {Casini}, {Centeno}, {Cranmer},
  {Criscuoli}, {Deforest}, {Deng}, {Erd{\'e}lyi}, {Fedun}, {Fischer},
  {Gonz{\'a}lez Manrique}, {Hahn}, {Harra}, {Henriques}, {Hurlburt}, {Jaeggli},
  {Jafarzadeh}, {Jain}, {Jefferies}, {Keys}, {Kowalski}, {Kuckein}, {Kuhn},
  {Kuridze}, {Liu}, {Liu}, {Longcope}, {Mathioudakis}, {McAteer}, {McIntosh},
  {McKenzie}, {Miralles}, {Morton}, {Muglach}, {Nelson}, {Panesar}, {Parenti},
  {Parnell}, {Poduval}, {Reardon}, {Reep}, {Schad}, {Schmit}, {Sharma},
  {Socas-Navarro}, {Srivastava}, {Sterling}, {Suematsu}, {Tarr}, {Tiwari},
  {Tritschler}, {Verth}, {Vourlidas}, {Wang}, {Wang}, {NSO and DKIST Project},
  {DKIST Instrument Scientists}, {DKIST Science Working Group}, \& {DKIST
  Critical Science Plan Community}}]{2021SoPh..296...70R}
{Rast}, M.~P., {Bello Gonz{\'a}lez}, N., {Bellot Rubio}, L., {et~al.} 2021,
  \solphys, 296, 70, \dodoi{10.1007/s11207-021-01789-2}

\bibitem[{{Ruzmaikin} \& {Berger}(1998)}]{1998A&A...337L...9R}
{Ruzmaikin}, A., \& {Berger}, M.~A. 1998, \aap, 337, L9

\bibitem[{{Sen} \& {Mangalam}(2019)}]{2019ApJ...877..127S}
{Sen}, S., \& {Mangalam}, A. 2019, \apj, 877, 127,
  \dodoi{10.3847/1538-4357/ab141a}

\bibitem[{{Srivastava} {et~al.}(2017){Srivastava}, {Shetye}, {Murawski},
  {Doyle}, {Stangalini}, {Scullion}, {Ray}, {W{\'o}jcik}, \&
  {Dwivedi}}]{2017NatSR...743147S}
{Srivastava}, A.~K., {Shetye}, J., {Murawski}, K., {et~al.} 2017, Scientific
  Reports, 7, 43147, \dodoi{10.1038/srep43147}

\bibitem[{{Thalmann} {et~al.}(2013){Thalmann}, {Tiwari}, \&
  {Wiegelmann}}]{2013ApJ...769...59T}
{Thalmann}, J.~K., {Tiwari}, S.~K., \& {Wiegelmann}, T. 2013, \apj, 769, 59,
  \dodoi{10.1088/0004-637X/769/1/59}

\bibitem[{{Thurgood} {et~al.}(2014){Thurgood}, {Morton}, \&
  {McLaughlin}}]{2014ApJ...790L...2T}
{Thurgood}, J.~O., {Morton}, R.~J., \& {McLaughlin}, J.~A. 2014, \apjl, 790,
  L2, \dodoi{10.1088/2041-8205/790/1/L2}

\bibitem[{{Tomczyk} \& {McIntosh}(2009)}]{2009ApJ...697.1384T}
{Tomczyk}, S., \& {McIntosh}, S.~W. 2009, \apj, 697, 1384,
  \dodoi{10.1088/0004-637X/697/2/1384}

\bibitem[{{Tomczyk} {et~al.}(2007){Tomczyk}, {McIntosh}, {Keil}, {Judge},
  {Schad}, {Seeley}, \& {Edmondson}}]{2007Sci...317.1192T}
{Tomczyk}, S., {McIntosh}, S.~W., {Keil}, S.~L., {et~al.} 2007, Science, 317,
  1192, \dodoi{10.1126/science.1143304}

\bibitem[{{Tomczyk} {et~al.}(2008){Tomczyk}, {Card}, {Darnell}, {Elmore},
  {Lull}, {Nelson}, {Streander}, {Burkepile}, {Casini}, \&
  {Judge}}]{2008SoPh..247..411T}
{Tomczyk}, S., {Card}, G.~L., {Darnell}, T., {et~al.} 2008, \solphys, 247, 411,
  \dodoi{10.1007/s11207-007-9103-6}

\bibitem[{{Tripathi} {et~al.}(2010){Tripathi}, {Mason}, {Del Zanna}, \&
  {Young}}]{2010A&A...518A..42T}
{Tripathi}, D., {Mason}, H.~E., {Del Zanna}, G., \& {Young}, P.~R. 2010, \aap,
  518, A42, \dodoi{10.1051/0004-6361/200913883}

\bibitem[{{Tripathi} {et~al.}(2009){Tripathi}, {Mason}, {Dwivedi}, {del Zanna},
  \& {Young}}]{2009ApJ...694.1256T}
{Tripathi}, D., {Mason}, H.~E., {Dwivedi}, B.~N., {del Zanna}, G., \& {Young},
  P.~R. 2009, \apj, 694, 1256, \dodoi{10.1088/0004-637X/694/2/1256}

\bibitem[{{van Ballegooijen} {et~al.}(2011){van Ballegooijen}, {Asgari-Targhi},
  {Cranmer}, \& {DeLuca}}]{2011ApJ...736....3V}
{van Ballegooijen}, A.~A., {Asgari-Targhi}, M., {Cranmer}, S.~R., \& {DeLuca},
  E.~E. 2011, \apj, 736, 3, \dodoi{10.1088/0004-637X/736/1/3}

\bibitem[{{van der Holst} {et~al.}(2014){van der Holst}, {Sokolov}, {Meng},
  {Jin}, {Manchester}, {T{\'o}th}, \& {Gombosi}}]{2014ApJ...782...81V}
{van der Holst}, B., {Sokolov}, I.~V., {Meng}, X., {et~al.} 2014, \apj, 782,
  81, \dodoi{10.1088/0004-637X/782/2/81}

\bibitem[{{Van Doorsselaere} {et~al.}(2016){Van Doorsselaere}, {Antolin},
  {Yuan}, {Reznikova}, \& {Magyar}}]{2016FrASS...3....4V}
{Van Doorsselaere}, T., {Antolin}, P., {Yuan}, D., {Reznikova}, V., \&
  {Magyar}, N. 2016, Frontiers in Astronomy and Space Sciences, 3, 4,
  \dodoi{10.3389/fspas.2016.00004}

\bibitem[{{Van Doorsselaere} {et~al.}(2014){Van Doorsselaere}, {Gijsen},
  {Andries}, \& {Verth}}]{2014ApJ...795...18V}
{Van Doorsselaere}, T., {Gijsen}, S.~E., {Andries}, J., \& {Verth}, G. 2014,
  \apj, 795, 18, \dodoi{10.1088/0004-637X/795/1/18}

\bibitem[{{Van Doorsselaere} {et~al.}(2020){Van Doorsselaere}, {Li},
  {Goossens}, {Hnat}, \& {Magyar}}]{2020ApJ...899..100V}
{Van Doorsselaere}, T., {Li}, B., {Goossens}, M., {Hnat}, B., \& {Magyar}, N.
  2020, \apj, 899, 100, \dodoi{10.3847/1538-4357/aba0b8}

\bibitem[{{Verbeeck} {et~al.}(2014){Verbeeck}, {Delouille}, {Mampaey}, \& {De
  Visscher}}]{2014A&A...561A..29V}
{Verbeeck}, C., {Delouille}, V., {Mampaey}, B., \& {De Visscher}, R. 2014,
  \aap, 561, A29, \dodoi{10.1051/0004-6361/201321243}

\bibitem[{{Warren} {et~al.}(2008){Warren}, {Ugarte-Urra}, {Doschek}, {Brooks},
  \& {Williams}}]{2008ApJ...686L.131W}
{Warren}, H.~P., {Ugarte-Urra}, I., {Doschek}, G.~A., {Brooks}, D.~H., \&
  {Williams}, D.~R. 2008, \apjl, 686, L131, \dodoi{10.1086/592960}

\bibitem[{{Young} {et~al.}(2012){Young}, {O'Dwyer}, \&
  {Mason}}]{2012ApJ...744...14Y}
{Young}, P.~R., {O'Dwyer}, B., \& {Mason}, H.~E. 2012, \apj, 744, 14,
  \dodoi{10.1088/0004-637X/744/1/14}

\end{thebibliography}
\bibliographystyle{aasjournal}

\end{document}